\documentclass[11pt,a4paper]{article}
\usepackage{amsfonts,amssymb,graphicx,color}
\usepackage{exscale}
\usepackage[sumlimits,intlimits]{amsmath}
\begin{document}
%\noindent {KTH-HH-6/2003 Rila-Workshop summer 2003}
%\\[0.5cm]
\begin{center}{\Large \bf
Relativistic Nucleon-Nucleon potentials using \\[0.1cm] 
Dirac's constraint instant form dynamics}
\\[0.5cm]
H.V. von Geramb\footnote{
Presented at XXII International Workshop
on Nuclear Theory, Rila Mountains, June 16-22 2003,\\ E-mail: geramb@uni-hamburg.de},
Davaadorj Bayansan and St. Wirsching 
\\[0.1cm]
{\em Theoretische Kernphysik, Universit\"at Hamburg
\\ Luruper Chaussee 149, D-22761 Hamburg}
\end{center}
\begin{abstract}
The formalism of two coupled  Dirac equations within 
constraint instant form dynamics is used to study the
nucleon-nucleon (NN) interaction. The salient features 
and the final Schr\"odinger type  equation is given. 
Explicitly energy dependent coupled channel potentials,
for use in partial wave Schr\"odinger like equations, with nonlinear 
and complicated derivative terms, result. We developed the necessary
numerics and study $np$ and $pp$ scattering phase shifts for
energies 0--3 GeV and the deuteron bound state. The interactions are
inspired by meson exchange of $\pi,\eta,\rho,\omega$ and
$\sigma$ mesons for which we adjust coupling constants. 
This yields, in the first instant, high quality fits to the Arndt phase
shifts 0--300 MeV. Second, the potentials show a universal, independent from
angular momentum, core potential which is generated from the 
relativistic meson exchange dynamics. Extrapolations towards higher energies,
up to $T_{Lab}=3$\,GeV, allow to separate a QCD dominated short range
zone as well as inelastic nucleon excitation mechanism  contributing to
meson production. A local and/or nonlocal  optical model, in addition to the
meson exchange Dirac potential, produces agreement between theoretical 
and data phase shifts. Third, the $^1S_0,\ ^3P_0$ and $ ^3P_1$ partial waves
elicit a fusion/scission, for $T_{Lab}<1$\,GeV, and a
fusion/fission, for $T_{Lab}>1$\,GeV, mechanism for intermediate
dibaryon formation. 
\end{abstract}
\section{Introduction}
The formalism of coupled two-body Dirac equations, within 
constraint instant form dynamics, is used to study the
nucleon-nucleon (NN) interaction. This particular approach for two spin
1/2 particles was developed by Crater, Van Alstine, Long and Liu 
\cite{Cra83,Cra87,Lon98,Cra02,Liu03,Cra03}.
They define a Poincar\'e invariant interaction in terms
of scalar, pseudo scalar, vector etc. interactions with the implication
that they satisfy certain compatibility conditions \cite{Lon98}.
This approach yields in its final form  explicitly energy dependent 
coupled channel potentials for use in partial wave Schr\"odinger 
like equations \cite{Liu03}. We followed and re-derived their expressions
up to a certain point and
developed our own numerics to study  $np$ and $pp$ 
scattering phase shifts for $0<T_{Lab}<3$\,GeV.
The comparison with recent data makes use of
SM00 and SP03 GWU/VPI SAID phase shift solutions \cite{GWU03}.

The NN interaction is described within the paradigm of exchange
mechanism involving $\pi,\rho,\omega\,\sigma$ and other mesons exchanges \cite{Mac01} 
to make up what we call the Dirac potential. 
A comparison with most recent experimental data, GWU/VPI SP03 phase shifts,
requires the adjustment of coupling constants and a regularization
of the short range interaction domain. For energies above pion production threshold 
$280< T_{Lab}< 3000$\,MeV we added a phenomenological 
complex optical model potential  to 
the Dirac potential. This addition brings the theoretical S-matrix
in perfect agreement with the experimental data S-matrix from GWU/VPI
and, more importantly, permits the identification of some predominant 
reaction paths.

The coupled two-body Dirac equations, combined with the meson exchange model,
yield the appearance of a repulsive, practically  
hard core, potential independent of partial wave. 
The universal core radius has a value $r_c=0.5\pm0.025$\,fm. This core
radius is independent of a nucleon substructure. It depends only on
masses, in particular of the exchanged mesons, 
and the full relativistic treatment of the NN system.  
This feature is not present with equal distinctness in any of the 
current NN best fit potentials
of {\em np} and {\em pp} data \cite{Sto93,Wir95,Mac01}. 
For purpose of comparison, we show results of the Argonne AV18 potential \cite{Wir95}. 

The fitting process of coupling
constants uses data in the sub-meson-production domain,
$0< T_{Lab}< 280$\,MeV, of {\em np} and {\em pp} partial wave phase shifts.
For $T_{Lab}>280$\,MeV, single and double
intrinsic nucleon excitations, $\Delta(3,3)$ and other low excited hadrons, 
as well as simple and complex reactive meson productions contribute. 
This is well known and demands beyond NN a more complex coupled channels
problem to solve. We curtail the problem to NN scattering 
using an optical model potential (OMP) addition \cite{Fun01,Ger01}. 
Despite  of complicated inelasticities, selection rules of 
angular momentum, isospin selection  and  the complex energy
dependences  
some of the partial waves show that the {\em real phase shifts} $\delta(T)$, 
are very well reproduced (extrapolated)  by the Dirac potential alone. 
Most clearly, this is realized in the
$^1S_0$,$^3P_0$ and  $^3P_1$ channels and $T_{Lab}< 1100$\,MeV. 
We have realized this fact before \cite{Fun01,Ger01} but wish now to 
support more convincingly the case of an intermediate
fusion/scission, for $T_{Lab}<1$\,GeV, and a fusion/fission, for $T_{Lab}>1$\,GeV,
 mechanism in which  two nucleons change briefly  into a 
compact dibaryon with subsequent  decay back into two nucleons and mesons. 

In Fig. 1 we show an intuitive and guiding scheme
which distinguishes  interaction domains as
function of separation between the two nucleons. 
\begin{figure}[h]
\begin{center}
\includegraphics[scale=0.38]{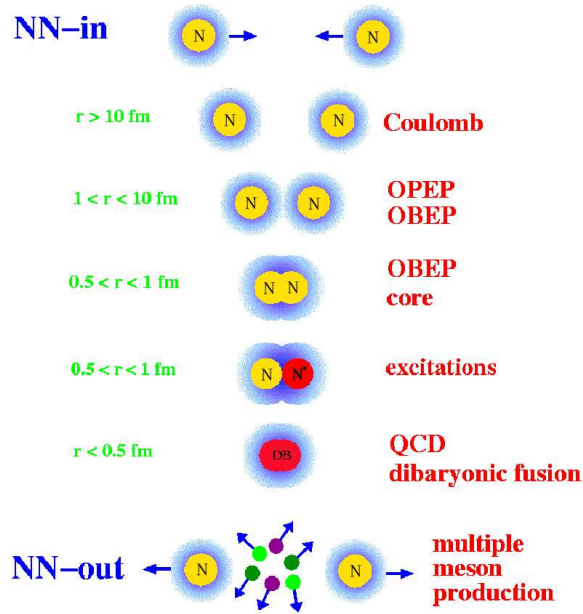}
\caption{NN scattering and reaction scheme for $T_{Lab}<3$\,GeV.}
\label{Figure_1}
\end{center}
\end{figure}
This scheme is in accordance with coupled channels. 

\section{Theoretical background}

Relativistic quantum mechanics demands a Poincar\'e invariant formulation. 
This implies unitary representations of the Poincar\'e group as 
transformations for state vectors. One distinguishes  
ten generators for translations and rotations
\begin{equation}
P_\mu (4)\quad\text{and}\quad J_{\mu\nu}(6)=-J_{\nu\mu}.
\end{equation}
The associated commutation relations yield the  Poincar\'e algebra.
Subsets of these generators  are associated with subgroups, which
correspond to 3D  hypersurfaces in Minkowski space. 
The kinematic subgroups are 
instant form $x^0 = 0$, light  front form  $x^0 + x^3 = 0$ 
and point form $x\cdot x = a^2 > 0$ with $x^0 > 0$ \cite{Dir49}. 
Within the fundamental variables, one distinguishes
simple  kinematic  variables and  complex Hamiltonians.

Crater, Van Alstine and their collaborators \cite{Cra83,Cra87,Lon98,Cra03,Liu03}
treat the {\em   nucleon-nucleon  problem}
with two coupled Dirac equations. Each of the two free nucleons satisfies 
\begin{equation}(i\gamma^\mu\partial_\mu-m)\psi=0\quad\text{and}\quad 
(\hat p^2+m^2)\psi=0.
\end{equation}
They relate to a representation in terms of Todorov's variables
\begin{gather*}
x=x_1-x_2\quad \text{relative distance}
\nonumber \\  
P=p_1+p_2\quad \text{total momentum}
\nonumber \\ 
\omega=\sqrt{-P^2}\quad \text{invariant mass, in C.M. total energy}
\nonumber \\
(P^2+\omega^2)\Psi=0,\quad\vec P=0\quad \text{in C.M.}
\nonumber \\
\hat P_\nu= \frac{P_\nu}{\omega},\quad x_\perp^\nu=(\eta^{\nu\mu}+\hat P^\nu 
\hat P^\mu)x_\mu\quad  \text{transverse coordinate}
\nonumber \\
\epsilon_1=\frac{\omega^2+m_1^2-m_2^2}{2\omega},\quad \epsilon_2=\frac{\omega^2+m_2^2-m_1^2}{2\omega}
\nonumber \\
m_\omega=\frac{m_1\,m_2}{\omega}\ \text{relativistic reduced mass}\nonumber \\ 
\epsilon_\omega=\frac{\omega^2-m_1^2-m_2^2}{2\omega}\quad \text{relativistic relative energy on
reduced mass}
\nonumber \\
p=\frac{\epsilon_2p_1-\epsilon_1p_2}{\omega},\quad P\cdot p=0,
\quad p_1=\frac{\epsilon_1 P}{\omega}+p,\quad
p_2=\frac{\epsilon_2 P}{\omega}-p
\nonumber \\
k^2=p^2=b^2(\omega,m_1,m_2)=
\frac{\omega^4+(m_1^2-m_2^2)^2-2\omega^2(m_1^2+m_2^2)}{ 4 \omega^2}\nonumber \\
=\epsilon^2_\omega-m_\omega^2=\epsilon_1^2-m_1^2=\epsilon_2^2-m_2^2. 
\end{gather*}
In particular $\theta$-matrices are related to $\gamma$-matrices. 
\begin{gather*}
\eta^{\mu\nu}=\eta_{\mu\nu}=\text{diag(-1,1,1,1)},\quad 
\theta ^\mu=i\sqrt{\frac12}\gamma_5\gamma^{\mu},\quad 
\theta _5=i\sqrt{\frac12}\gamma_5,\\
\theta_\perp^\nu=(\eta^{\nu\mu}-P^\nu P^\mu/P^2)\theta_\mu
\end{gather*}
Several anticommutator relations hold, {\em viz.} 
\begin{gather*}
[\theta ^\mu,\theta ^\nu]_{+}=-\eta^{\mu\nu}, \quad
[\theta _5,\theta _5]_{+}=-1,\\ [\theta _5,\theta ^\mu]_{+}=0,\quad
[\theta_i\cdot \hat P,\theta_i\cdot \hat P]_+=1,\quad 
[\theta_i\cdot\hat P,\theta_{\perp\, i}^\mu]_+=0.
\end{gather*}
The free equations are 
\begin{equation}
( p\cdot  \theta +m \theta _5)\,\psi=0\quad\text{and}\quad
(p^2+m^2)\psi=0.
\end{equation}
The generators of the  Lorentz group $J_{\mu\nu}$ contain angular momentum and
spin
\begin{gather*}
J_{\mu\nu}=L_{\mu\nu}+S_{\mu\nu}\\ 
L_{\mu\nu}=\frac{1}{2}(q_\mu p_\nu+p_\nu
q_\mu-q_\nu p_\mu- p_\mu q_\nu)\\
 S_{\mu\nu}=-\frac{i}{2}( \theta _\mu \theta _\nu-
\theta _\nu \theta _\mu)
=\frac{i}{4}(\gamma_\mu\gamma_\nu-\gamma_\nu\gamma_\mu).
\end{gather*}
Each nucleon moves in the field of the other nucleon.
Ultimately, the interaction is described by a meson exchange model.

For two  particles with spin       
$${\cal S}_{i0}=p_i\theta_i+m_i\theta_{5i}, \quad i=1,2$$  
the commutator $$[{\cal S}_{10},{\cal S}_{20}]=0$$ vanishes strongly.

\subsection{Example with scalar interaction} 

A scalar interaction  changes the mass into a mass operator
$M_i=m_i+S$. The  Dirac equations are  
\begin{gather*}
{\mathcal S}_1\psi=(+p\cdot\theta _1+\epsilon_1\hat P \cdot\theta_1+M_1\theta _{51})\psi=0\nonumber \\
{\mathcal S}_2\psi=(-p\cdot\theta_2+\epsilon_2\hat P\cdot\theta_2+M_2\theta _{52})\psi=0\nonumber 
\\[0.1cm]
[\mathcal{S}_1,\mathcal{S}_2]\_\,\psi=-i\,(\partial M_1\cdot\theta_1\theta _{52}
+\partial M_2\cdot\theta_2\theta _{51})\psi\ne 0.
\end{gather*} 
The commutator does not vanish in general but vanishes through a {\em third law} condition 
\begin{equation}
\partial (M_1^2(x_\perp)-M_2^2(x_\perp))=0\quad\text{and}\quad
M_1^2(x_\perp)-M_2^2(x_\perp)=m_1^2-m_2^2.
\end{equation}
Using a  hyperbolic parameterization 
\begin{gather*}
M_1=m_1 \cosh L+m_2 \sinh L\nonumber \\
M_2=m_2 \cosh L+m_1 \sinh L
\end{gather*}
gives, in terms of a single invariant function $L=L(x_\perp)$,  
a compatible representation of the scalar interaction. 
The two body Dirac equations, with a scalar interaction, are
\begin{gather*}
\mathcal{S}_1\psi=\left(+ \theta_1\cdot p+\epsilon_1\theta_1\cdot
\hat P+M_1\theta_{51}-i\partial L\cdot\theta_2\theta_{52}\theta_{51}\right)\psi=0\nonumber \\
\mathcal{S}_2\psi=\left(-\theta_2\cdot p+\epsilon_2\theta_2\cdot
\hat P+M_2\theta_{52}+i\partial L\cdot\theta_1\theta_{51}\theta_{52}\right)\psi=0\nonumber \\
\text{where}\quad\quad \partial L=\frac{\partial M_1}{M_2}=\frac{\partial M_2}{M_1}.
\end{gather*}

The Dirac constraint operators satisfy
\begin{equation}
\left(\mathcal{S}_1^2- \mathcal{S}_2^2\right)\psi=
-\frac12\left(p_1^2+m_1^2-p_2^2-m_2^2\right)\psi=-P\cdot p\,\psi=0.
\end{equation}

Crater, Van Alstine and collaborators generalized their 
hyperbolic representation to apply for any interaction being Poincar\'e invariant, {\em viz.} 
\begin{description}
\item{Scalar:}
$$ 
\Delta_L=-L\,\theta_{51}\theta_{52}=-\frac{L}{2}\mathcal{O}_1\quad\text{with}\quad
\mathcal{O}_1=-\gamma_{51}\gamma_{52}
$$ 
\item{Time-like Vector:}
$$
\Delta_J=J\,\hat P \cdot\theta_1 \hat P\cdot
\theta_2=\mathcal{O}_2\frac{J}{2}=\beta_1\beta_2\frac{J}{2}\mathcal{O}_1
$$
\item{Space-like Vector:}
$$
\Delta_{\mathcal G}=\mathcal{G}\,\theta_{1\perp}\cdot\theta_{2\perp}
=\mathcal{O}_3\frac{\mathcal{G}}{2}=
\gamma_{1\perp}\gamma_{2\perp}\frac{\mathcal{G}}{2}\mathcal{O}_1
$$
\item{Pseudo-scalar:}
$$
\Delta_C=\mathcal{E}_1\frac{C}{2}=-\gamma_{51}\gamma_{52}\mathcal{O}_1\frac{C}{2}
$$
\end{description}
and four others of which we shall not make any use. 
With such  sum of interactions
\begin{equation}
\Delta=\Delta_{J}+\Delta_{L}+\Delta_{\mathcal{G}}+\Delta_{C},
\end{equation}
the coupled system of equations 
\begin{gather*}
\mathcal{S}_{1}\psi=\left(+ G\theta_{1}\cdot
p+E_{1}\theta_{1}\cdot\hat P +M_{1}\theta_{51}+i \frac{G}{2}\left(\theta_{2}
\cdot\partial \mathcal{G}\mathcal{O}_{3}+J\mathcal{O}_{2}-
L\mathcal{O}_{1}+C\mathcal{E}_{1}\right)\right)\psi=0,
\nonumber\\[0.1cm]
\mathcal{S}_{2}\psi=\left(-G\theta_{2}\cdot p
+E_{2}\theta_{2}\cdot\hat{P}+M_{2}\theta_{52}-i \frac{G}{2}\left(\theta_{1}
\cdot\partial \mathcal{G}\mathcal{O}_{3}+J\mathcal{O}_{2}-L\mathcal{O}
_{1}+C\mathcal{E}_{1}\right)\right)\psi=0  
\end{gather*}
are fully defined by masses, energies and single particle quantities 
\begin{gather*}
M_{1}=m_{1}\cosh(L)+m_{2}\sinh(L)
\nonumber \\
M_{2}=m_{2}\cosh(L)+m_{1}\sinh(L)\\[0.1cm]
E_{1}=\epsilon_{1}\cosh(J)+\epsilon_{2}\sinh(J)
\nonumber \\
E_{2}=\epsilon_{2}\cosh(J)+\epsilon_{1}\sinh(J)\label{E1E2_eqn}\\[0.1cm]
\text{and}\quad\quad G=e^{\mathcal G}.
\end{gather*}

Finally an elaborate Pauli reduction yields coupled
Schr\"odinger type equations.  
$\sigma,\,\rho,\,\omega,\,\pi$ and $\eta$ exchanges  specify the interactions
\begin{equation}
L(x_{\perp}),J(x_{\perp}),C(x_{\perp}),\mathcal{G}(x_{\perp}).
\end{equation}
We use
\begin{equation}
\mathcal{D}= E_1M_2+E_2M_1
\end{equation}
and
\begin{equation}
\mathcal{B}^2=E_1^2-M_1^2=E_2^2-M_2^2.
\end{equation}

The final stationary Schr\"odinger type equation  
\begin{gather}
\label{schroe_eqn}
\left\{
\mathbf{p}^2-i\left[ 2\mathcal{G}^{\prime}-\frac{E_2M_2+M_1E_1}
{\mathcal{D}}(J+L)^{\prime}\right]( {\mathbf{\hat r}}\cdot\mathbf{p})
\right. \nonumber \\ 
-\frac{i(J-L)^{\prime}}{2}( (\mathbf{\sigma}_1
\cdot{\mathbf{\hat r}}) (\mathbf{\sigma}_{2}\cdot\mathbf{p})+
(\mathbf{\sigma}_{2}\cdot{\mathbf{\hat r}})
(\mathbf{\sigma}_{1}\cdot\mathbf{p}))
\nonumber \\ 
-\frac{1}{2}\nabla^{2}\mathcal{G-}\frac{1}{4}\mathcal{G}^{\prime2}
-\frac{1}{4}(C+J-L)^{\prime}(-C+J-L)^{\prime}+\frac{1}{2}\frac{E_{2}
M_{2}+M_{1}E_{1}}{\mathcal{D}}\mathcal{G}^{\prime}(J+L)^{\prime}
\nonumber \\ 
+(\mathbf{\sigma}_{1}\cdot\mathbf{\sigma}_{2})\left[ \frac{1}{2}\nabla
^{2}\mathcal{G+}\frac{1}{2}\mathcal{G}^{\prime2}-\frac{1}{2}\frac{E_{2}
M_{2}+M_{1}E_{1}}{\mathcal{D}}\mathcal{G}^{\prime}(J+L)^{\prime}-\frac{1}
{2}\mathcal{G}^{\prime}C^{\prime}
-\frac{1}{2}\frac{\mathcal{G}^{\prime}}
{r}-\frac{1}{2}\frac{(-C+J-L)^{\prime}}{r}\right]
\nonumber \\ 
+\frac{\mathbf{L}\cdot(\mathbf{\sigma}_{1}+\mathbf{\sigma}_{2})}
{r}\left[ \mathcal{G}^{\prime}-\frac{1}{2}\frac{E_{2}M_{2}+M_{1}E_{1}}{\mathcal{D}
}(J+L)^{\prime}\right]
\nonumber \\ 
-\frac{\mathbf{L}\cdot(\mathbf{\sigma}_{1}-\mathbf{\sigma}_{2})}
{r}\frac{1}{2}\frac{E_{2}M_{2}-M_{1}E_{1}}{\mathcal{D}}(J+L)^{\prime}
\nonumber \\ 
+\frac{\mathbf{L}\cdot(\mathbf{\sigma}_{1}\times\mathbf{\sigma}_{2})}{r}\frac{i}
{2}\frac{M_{2}E_{1}-M_{1}E_{2}}{\mathcal{D}}(J+L)^{\prime}
\nonumber \\ 
 +(\mathbf{\sigma}_1\cdot{\mathbf{\hat r}})(\mathbf{\sigma}_2\cdot
{\mathbf{\hat r}})
\left[
 -\frac{1}{2}\nabla^{2}(-C+J-L)-\frac{1}{2}\nabla
^{2}\mathcal{G-G}^{\prime}(-C+J-L)^{\prime}-\mathcal{G}^{\prime2}+\frac{3}
{2r}\mathcal{G}^{\prime}
\right. 
\nonumber \\ 
\left.\left.
+\frac{3}{2r}(-C+J-L)^{\prime}+\frac{1}{2}\frac{E_{2}M_{2}+M_{1}E_{1}}
{\mathcal{D}}(J+L)^{\prime}(\mathcal{G}-C+J-L)^{\prime}
\right]
\right\}
|\phi_+\rangle
\nonumber  \\
=e^{-2\mathcal{G}}\mathcal{B}^2|\phi_{+}\rangle
\end{gather}
can be treated with well known techniques, in particular partial wave
expansion, to find angular momentum (with spin, isospin and angular momentum) 
dependent radial Schr\"odinger equations.

\section{Dirac Potentials and Partial Wave Phase Shifts}

Partial wave expansion separates  spin, angular and radial parts
to yield  coupled or uncoupled radial Schr\"odinger type equations 
\begin{equation}
-\phi''(r,k)+V_1(r)\phi'(r,k)+V_0(r)\phi(r,k)=k^2\phi(r,k).
\label{schroe_210_eqn}
\end{equation}
The appearance of a first derivative term requires solving a system of first
order equations which  is numerically not favorable. However, the Numerov algorithm 
is favorable and popular for radial Schr\"odinger equations
but applies only to second order, without first derivative, equations. 
A factorization of the solution $\phi(r,k)$ by the ansatz 
$$\phi(r,k):=g(r,k)f(r,k)$$ 
yields first order equations
\begin{equation}
g'(r,k)=\frac{1}{2}V_1(r) g(r,k)\quad\text{with}\quad \lim_{r \to \infty }g(r,k)=1  
\end{equation}
or second order equations
\begin{equation}
g''(r,k)=\left(\frac{1}{2}V_1'(r)+\frac{1}{4}V_1^2(r)\right)g(r,k)
\quad\text{with}\quad \lim_{r \to \infty }g(r,k)=1.  
\end{equation}
The other factor satifies the second order equation
\begin{equation}
-f''(r,k)+V(r)f(r,k)=k^2f(r,k), \quad  \lim_{r \to \infty }f(r,k)\sim 
\frac{1}{2i}(-h^{-}(r,k,\eta)+h^+(r,k,\eta)\,S(k)).
\end{equation}
Ricatti Hankel or Coulomb functions $h^\pm(r,k,\eta)$ and the
 S-matrix $S(k)$ determine the  asymptotic solutions.
We identify the  Dirac potential with
\begin{equation}
V^D(r)=g^{-1}(r,k)\left(\frac{1}{4}V_1^2(r)-\frac{1}{2}V_1'(r)+V_0(r)\right)g(r,k)
\end{equation} 
and additionally centrifugal and Coulomb potentials  
\begin{equation}
V^{FC}(r)=g^{-1}(r,k)\left( \frac{\ell(\ell+1)}{r^2}+\frac{2 k\,\eta(k)}{r}\right)g(r,k).
\end{equation}
We use the convention of \cite{Lon98,Liu03}
\begin{gather*}
P=p_1+p_2\quad \text{total momentum},\nonumber \\ 
\omega=\sqrt{-P^2}\quad \text{invariant mass, in C.M. total energy,}
\quad \vec P=0\quad \text{in C.M.},\nonumber \\
\epsilon_1=\frac{\omega^2+m_1^2-m_2^2}{2\omega},\quad \epsilon_2=\frac{\omega^2+m_2^2-m_1^2}{2\omega},
\nonumber \\
m_\omega=\frac{m_1\,m_2}{\omega}\quad \text{relativistic reduced mass},\nonumber \\ 
\epsilon_\omega=\frac{\omega^2-m_1^2-m_2^2}{2\omega}\quad \text{relativistic relative energy,}
\nonumber \\
p=\frac{\epsilon_2p_1-\epsilon_1p_2}{\omega},\quad p_1=\frac{\epsilon_1 P}{\omega}+p,\quad
p_2=\frac{\epsilon_2 P}{\omega}-p,\nonumber \\
k^2=p^2=
\frac{\omega^4+(m_1^2-m_2^2)^2-2\omega^2(m_1^2+m_2^2)}{ 4 \omega^2}=\epsilon^2_\omega-m_\omega^2
=\epsilon_1^2-m_1^2=\epsilon_2^2-m_2^2,\nonumber \\  
\eta(k)=\frac{\epsilon_\omega\,e^2}{k}\delta_{pp}\quad \text{Coulomb parameter.}\nonumber
\end{gather*}
This is to be compared with standard non-relativistic NN potentials \cite{Wir95}. The
recent work by Liu and Crater \cite{Liu03} elaborates on other methods to determine phase shifts
from Eq.\,(\ref{schroe_210_eqn}).

The model specification for 
$L(x_{\perp}),J(x_{\perp}),C(x_{\perp})$ and ${\mathcal G}(x_{\perp})$ follows
Liu and Crater {\em  model I} \cite{Liu03} to specify 
scalar  
$$
S=-g^2_\sigma\frac{e^{-m_\sigma r}}{r}-(\tau_1\cdot\tau_2)g^2_{a_0}
\frac{e^{-m_{a_0}r}}{r}-g^2_{f_0}\frac{e^{-m_{f_0}r}}{r},
$$
pseudo scalar
$$
C=(\tau_1\cdot\tau_2)\frac{g^2_{\pi}}{\omega}\frac{e^{-m_{\pi}r}}{r}+
\frac{g^2_{\eta}}{\omega}\frac{e^{-m_{\eta}r}}{r}-
\frac{g^2_{\eta'}}{\omega}\frac{e^{-m_{\eta '}r}}{r},
$$
and vector
$$
A=(\tau_1\cdot\tau_2)g^2_\rho\frac{e^{-m_{\rho}r}}{r}+g^2_\omega
\frac{e^{-m_{\omega}r}}{r}+g^2_\phi\frac{e^{-m_{\phi}r}}{r}
$$
interactions. All Yukawa form factors are regularized with a normalized  Gaussian
\begin{equation}
\frac{e^{-m\,r}}{r}\to N_G(a)\int dx^3\,\frac{e^{-m\,x}}{x}
\,e^{-(\vec r-\vec x)^2/a^2}\quad\text{with}\quad a=0.14142\,\text{fm.}
\end{equation}
For  $S<0$ we use
\begin{equation}
\sinh (L) =\frac{S\,G^2}{w}\left(1+\frac{G^2(\epsilon_\omega-A)S}{m_\omega\sqrt{\omega^2+S^2}}
\right),
\end{equation}
and  for $S>0$
\begin{gather*}
M_1^2=m_1^2+G^2(2m_wS+S^2)\\
M_2^2=m_2^2+G^2(2m_wS+S^2)\\
\partial L =\frac{\partial M_1}{M_2} =\frac{\partial M_2}{M_1}.
\end{gather*}
This models $\pi, \,\eta,\, \rho, \,\omega,\,\delta$ and $\sigma$ exchanges.
In Figs.\,\ref{Pot_1}, \ref{Pot_2}, \ref{Pot_3} and \ref{Pot_4} are shown Dirac potentials for
three values, $T_{Lab}=[0.1,1,2]$\,GeV (red lines). In comparison are
shown the results of the popular Argonne  AV18 potential (blue line) \cite{Wir95}. The remarkable feature
of the Dirac potentials is their universal repulsive core with $r_c\sim 0.5$\,fm.
The only exception is the $^3PF_1$ channel where the ansatz of repulsion turns, surprisingly,
into a short range attraction.

In Figs.\,\ref{Phase_1} and \ref{Phase_2} are shown the phase shifts of 
SM00 (green), SP03 (blue) and theoretical results 
(real Dirac potential solutions (red), real Dirac potentials
with complex OMP added are {\em coinciding} with the data of SP03 blue lines). 

The Dirac instant form dynamic yields partial wave  spin, isospin and energy ($\alpha$ channel)
 dependent NN  potentials $V_{\alpha}^D(r,T)$ to which  we add
a local or nonlocal optical model potential whose strengths are fitted to data
\begin{equation}  
\label{dirac_omp}\topmargin-1cm
V(r)=V_{\alpha}^D(r,T)+[U_\alpha(T) g(r)+iW_\alpha(T) g(r)]
\end{equation}
or
\begin{equation}
V(r,r')=V_{\alpha}^D(r,T)\delta(r-r')+
|\phi(r)>[U_\alpha(T)+iW_\alpha(T)]<\phi(r')|. 
\end{equation}
Since local/nonlocal potentials imply similar results, 
 we restrict ourselves here  to the
local optical potential and reference  the more general and
nonlocal case \cite{Fun01}.
  
A remark about phase shift convention appears necessary. 
Consider the partial wave radial equation 
\begin{equation}
-f''_\alpha(r,k)+V_\alpha(r)f_\alpha(r,k)=k^2f_\alpha(r,k)
\end{equation}
with stationary
\begin{equation}
f^{0}\sim j(kr)+n(kr)\,K(k)
\end{equation}
or physical  boundary conditions
\begin{equation}
\quad f^+\sim \frac1{2i}(-h^-(kr)+h^+(kr)\,S(k)).
\end{equation} 
$S$- and $K$-matrices are related to phase shifts \cite{Arn82}.
There exist several conventions to represent the S-matrix in terms of
phase shifts.  The Arndt and Roper \cite{Arn82} (GWU/VPI) convention 
uses   S- and K-matrices
\begin{equation}
S=(1+iK)(1-iK)^{-1},\quad\text{with}\quad K=i(1-S)(1+S)^{-1}
=\text{Re}\,K+i\,\text{Im}\, K.
\end{equation}
$\text{Re}\,K$ corresponds to a unitary S-matrix and phase shifts $\delta^\pm(k)$
and $\varepsilon(k)$ defined by
\begin{equation}S(\text{Re\,}K)=
\begin{pmatrix} \cos(2\varepsilon) \exp(2i\delta^-) &
  i\sin(2\varepsilon) \exp(i(\delta^- + \delta^+)) \\
 i\sin(2\varepsilon)\exp(i(\delta^- + \delta^+)) & \cos(2\varepsilon)
 \exp(2i\delta^-)
\end{pmatrix}.
\end{equation}
Absorption phase shifts, $\rho^\pm$ and $\mu$, are related to
\begin{equation}
\text{Im\,}K=\begin{pmatrix}\tan^2\rho^- & \tan\rho^-\tan\rho^+\cos\mu \\
 \tan\rho^-\tan\rho^+\cos\mu & \tan^2 \rho^+ \end{pmatrix}.
\end{equation}
Single channels simplify to $K=\tan\delta+i \tan^2\rho$. 

\section{The Nucleon-Nucleon Optical Potential}

The notion of an optical model is useful in cases when the S-matrix is
not unitary and flux disappears into open inelastic or reaction channels.
The optical model is often  expressed in terms of a complex and
energy dependent potential where the imaginary part effectively describes 
the loss of flux without specification of the inelastic channels. 
A less popular alternative to a complex  optical model potential is the 
introduction of pseudo channels. Here we follow the optical potential
approach \cite{Fun01}.

\begin{figure}%[h]
\begin{center}
\includegraphics[scale=0.55]{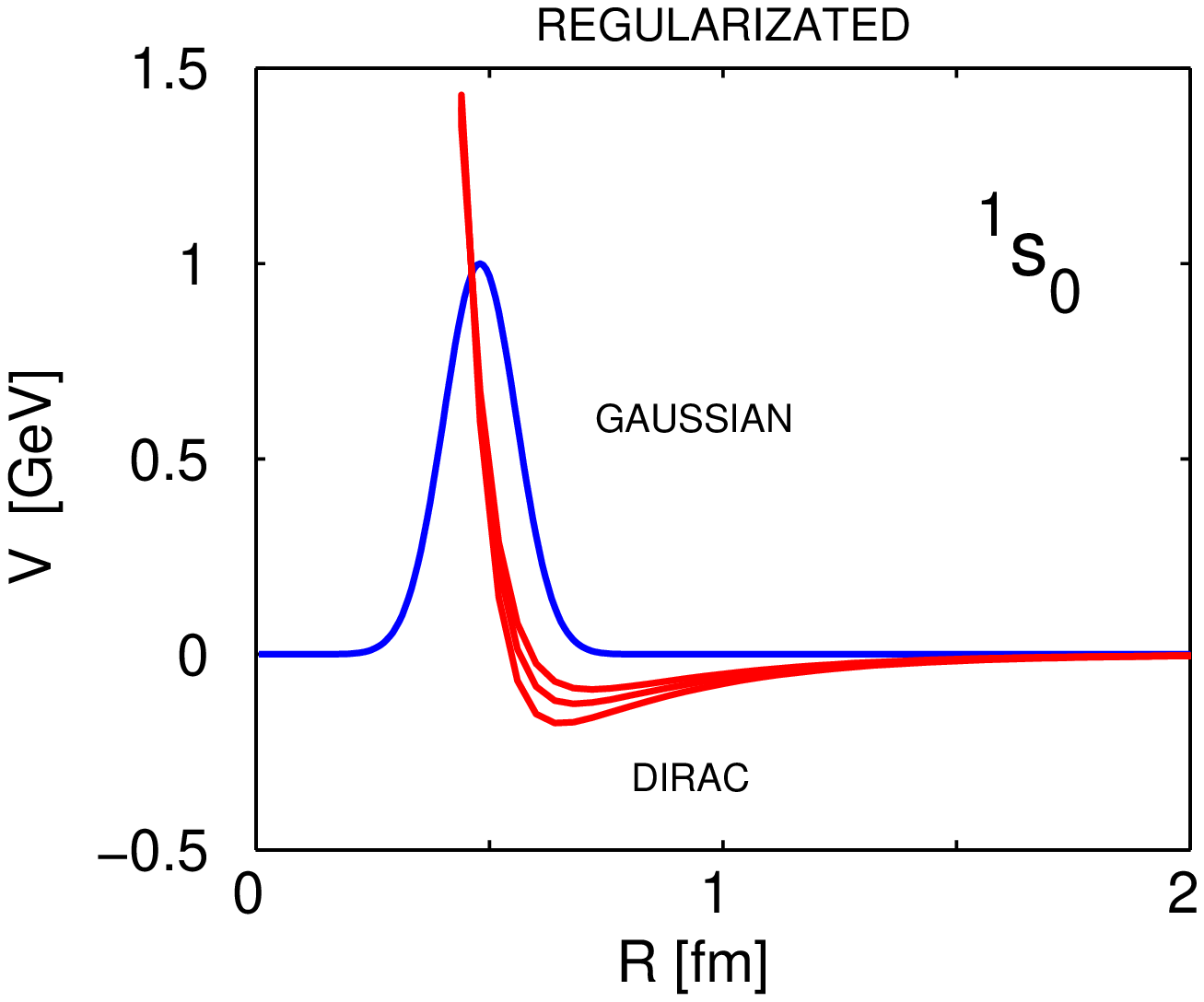}\quad
\includegraphics[scale=0.55]{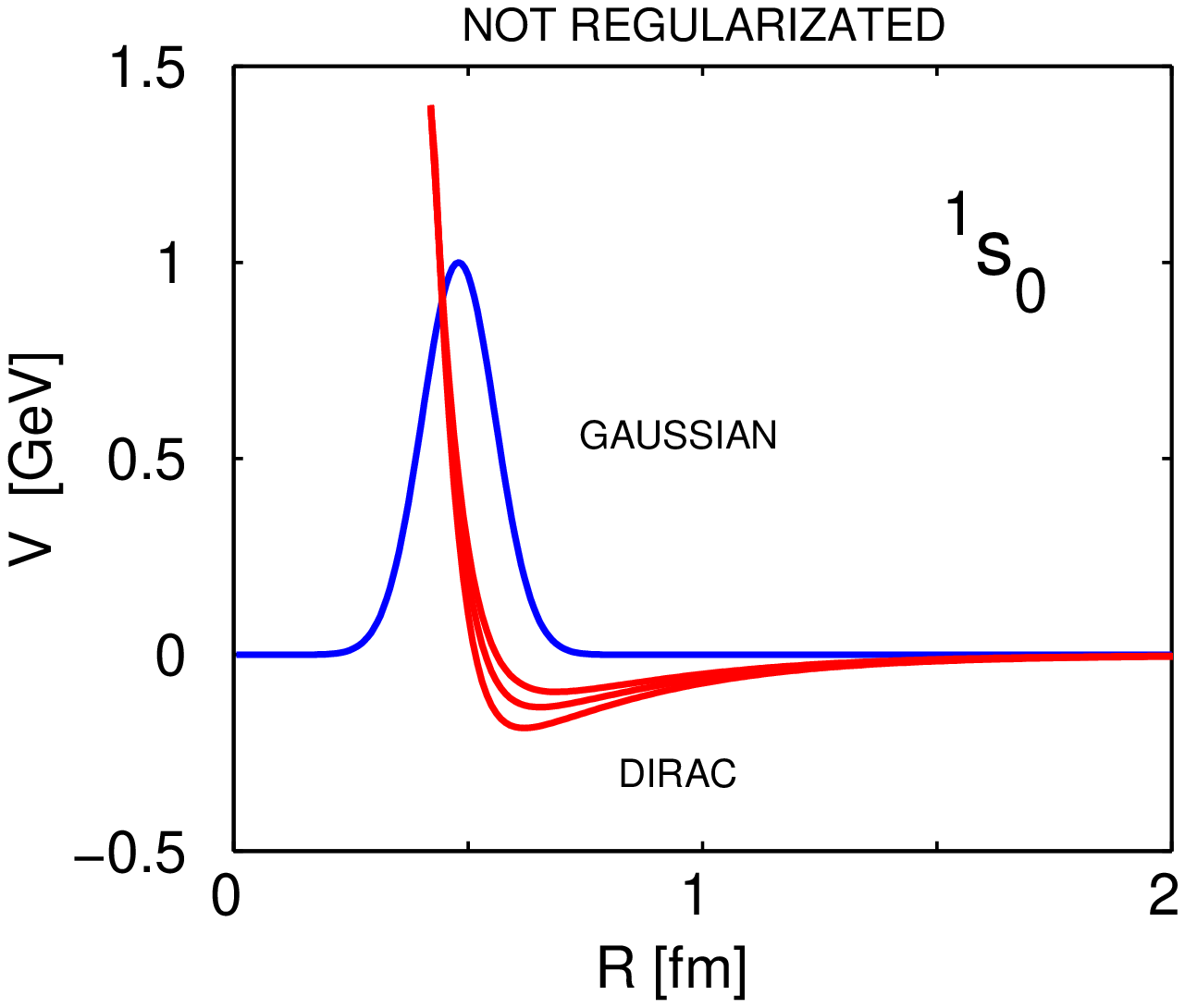}
\caption{
Dirac potential (red), left figure regularized and right figure not regularized, 
for the {\em np} $^1S_0$ channel, $T_{Lab}$ 0.1, 1 and 2 GeV,
showing a long range OPEP tail, an attractive pocket $\sim 0.75$\,fm and a core repulsion with
$r_c=0.5\pm 0.025$\,fm (blue). The regularized and not-regularized
potentials show, in all channels, only small differences. Regularization does not affect 
the conclusion drawn about the core geometry generated but helps to keep numbers reasonable near
the origins. Also inserted are   
Gaussian form factors $g(r-r_0)\sim\exp(-(r-r_0)^2/a^2)$ which are
used with the optical model (thin line curve). In this figure $r_0=r_c=0.5$\,fm  and $a=0.2$\,fm.}
\label{Figure_5}
\end{center}
\end{figure}

\begin{figure}[ht]
\begin{center}
\includegraphics[scale=1.0]{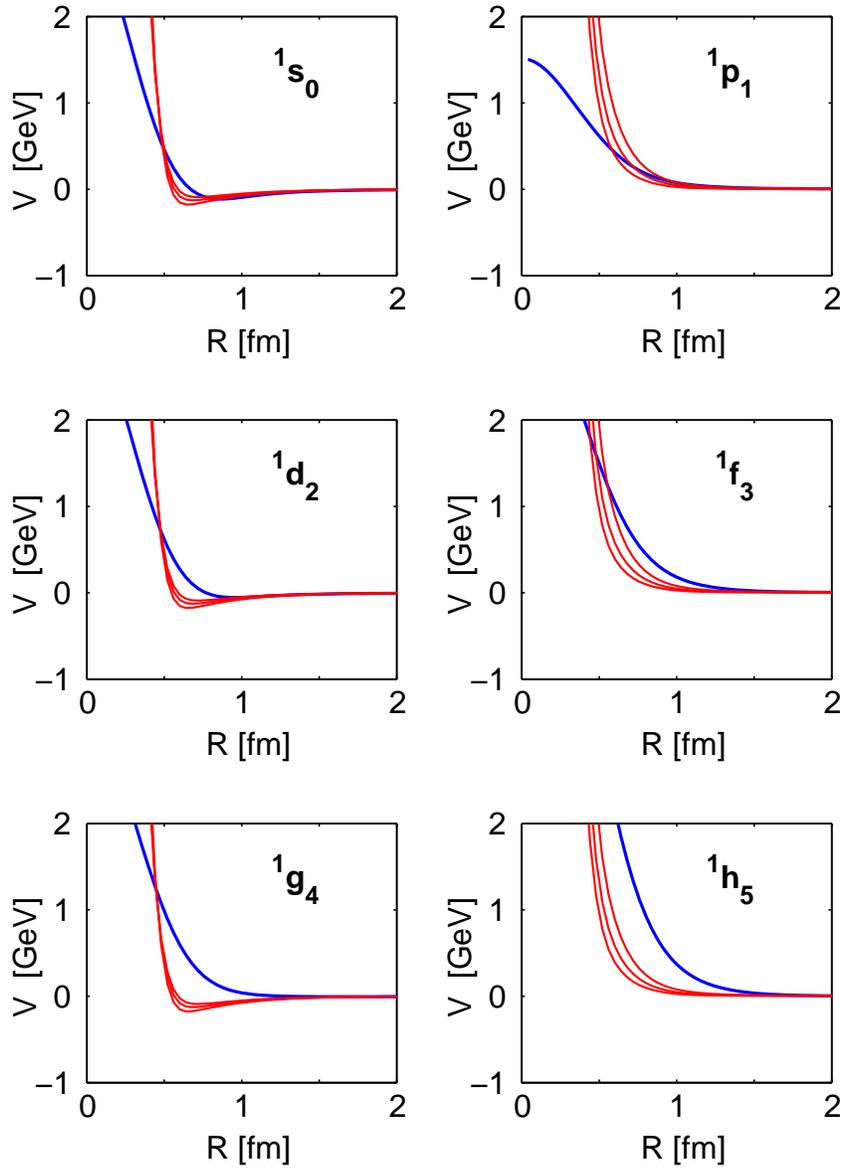}
\caption{Dirac potentials (red) for singlet $S=0$, $T=0$ (right column) and $T=1$ (left column) channels.
Differences between $np$ and $pp$ potentials are very small. The Coulomb potential is  added for
{\em pp}. The AV18 channel potential is shown as blue line. }\label{Pot_1}
\end{center}
\clearpage 
\end{figure}

\begin{figure}[ht]
\begin{center}
\includegraphics[scale=1.0]{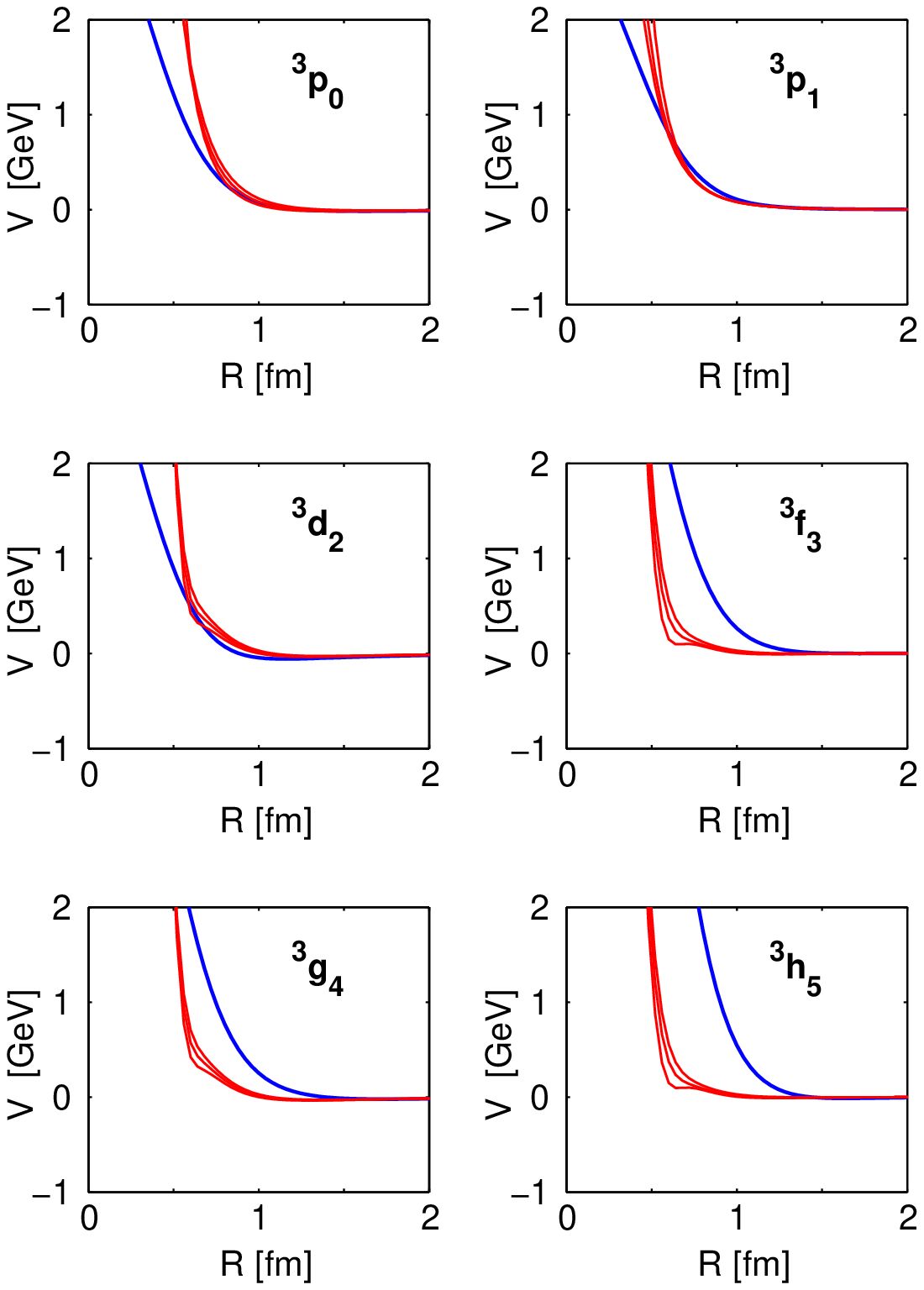}
\caption{Dirac potentials (red) for triplet $S=1$, $T=0,1$ uncoupled channels. 
The AV18 channel potential is shown as blue line.}\label{Pot_2}
\end{center}
\end{figure}

\begin{figure}[ht]
\begin{center}
\includegraphics[scale=1.0]{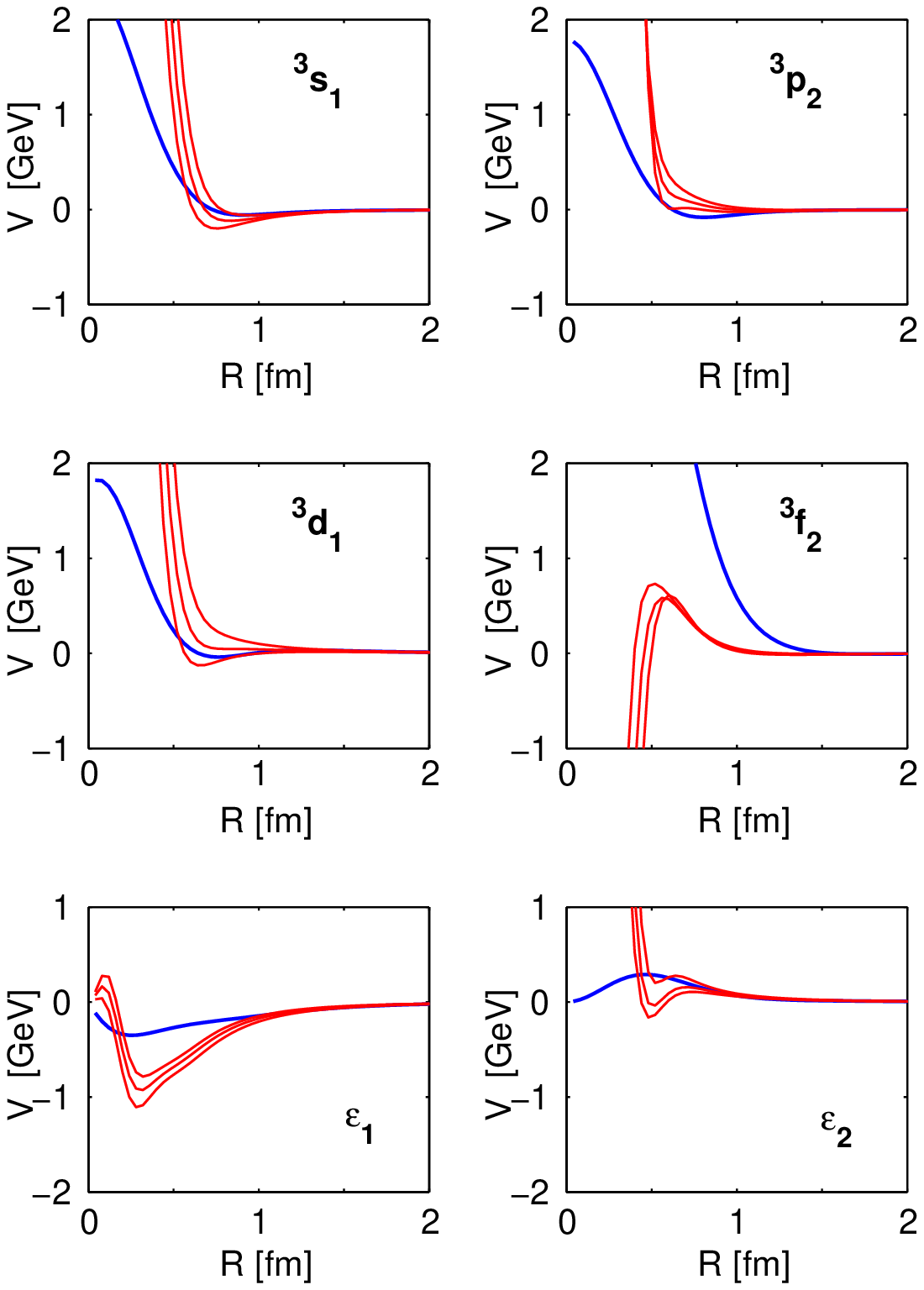}
\caption{ Dirac potentials (red) for triplet $S=1$, $T=0,1$ coupled channels.
 The AV18 channel potential is shown as blue line.}\label{Pot_3}
\end{center}
\end{figure}

\begin{figure}[ht]
\begin{center}
\includegraphics[scale=1.0]{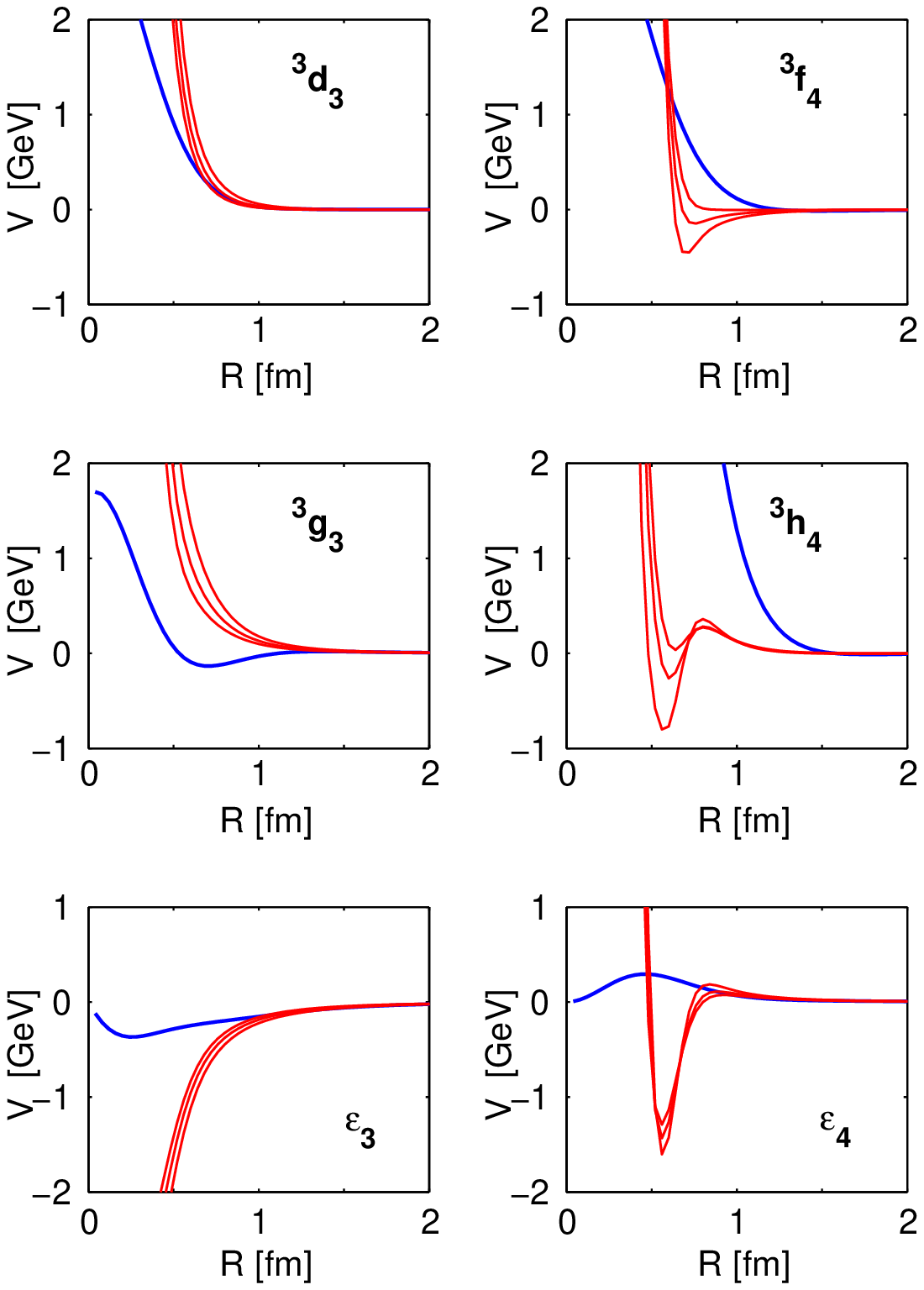}
\caption{ Dirac potentials (red) for triplet $S=1$, $T=0,1$ coupled channels.
 The AV18 channel potential is shown as blue line.}\label{Pot_4}
\end{center}
\end{figure}

\begin{figure}[ht]
\begin{center}
\includegraphics[scale=1.0]{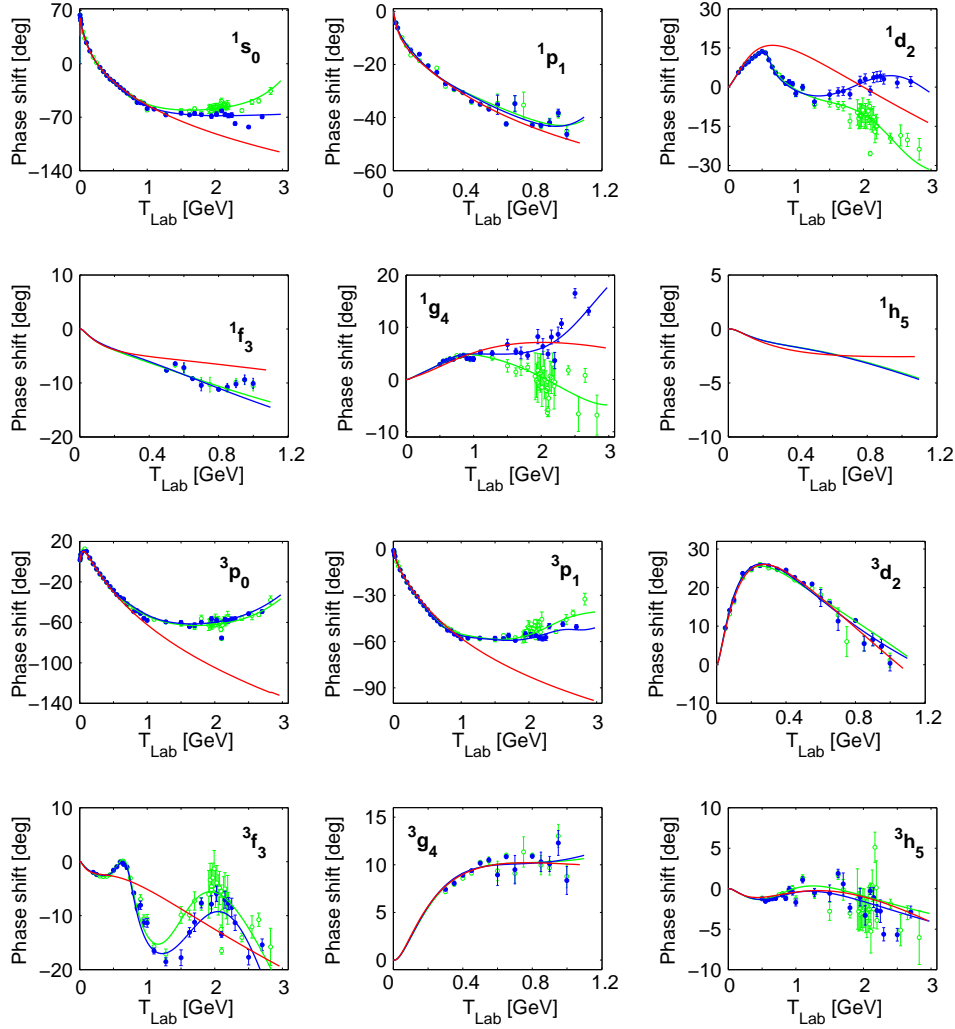}
\caption{Single channel,
$pp$ [0,3]\,GeV, $np$ [0,1.2]\,GeV SM00 (green) and SP03 (blue) real phase shift data $\delta(T)$ and
theoretical curve based soly upon Dirac potential (red).
The full potential, Dirac and OMP, describes the data SP03.  
Of particular interest are $^1S_0,^3P_0,^3P_1$ channels in which
the real phase shift of the Dirac potentials coincide with data for [0,1.1]\,GeV but diverge
above 1.1\,GeV. Obvious are effects due to $\Delta(3,3)$ coupling in other channels.
SM00 (continuous solution (green), single energy solutions, open green circles, with error bars). 
SP03 (continuous solution (blue), single energy solutions,full blue circles, with error bars).
Theoretical results (real Dirac potential solutions, full red line, real Dirac potentials
with complex OMP added are {\em coinciding} with the data of SP03 blue line).}
\label{Phase_1}
\end{center}
\end{figure}

\begin{figure}[ht]
\begin{center}
\includegraphics[scale=1.0]{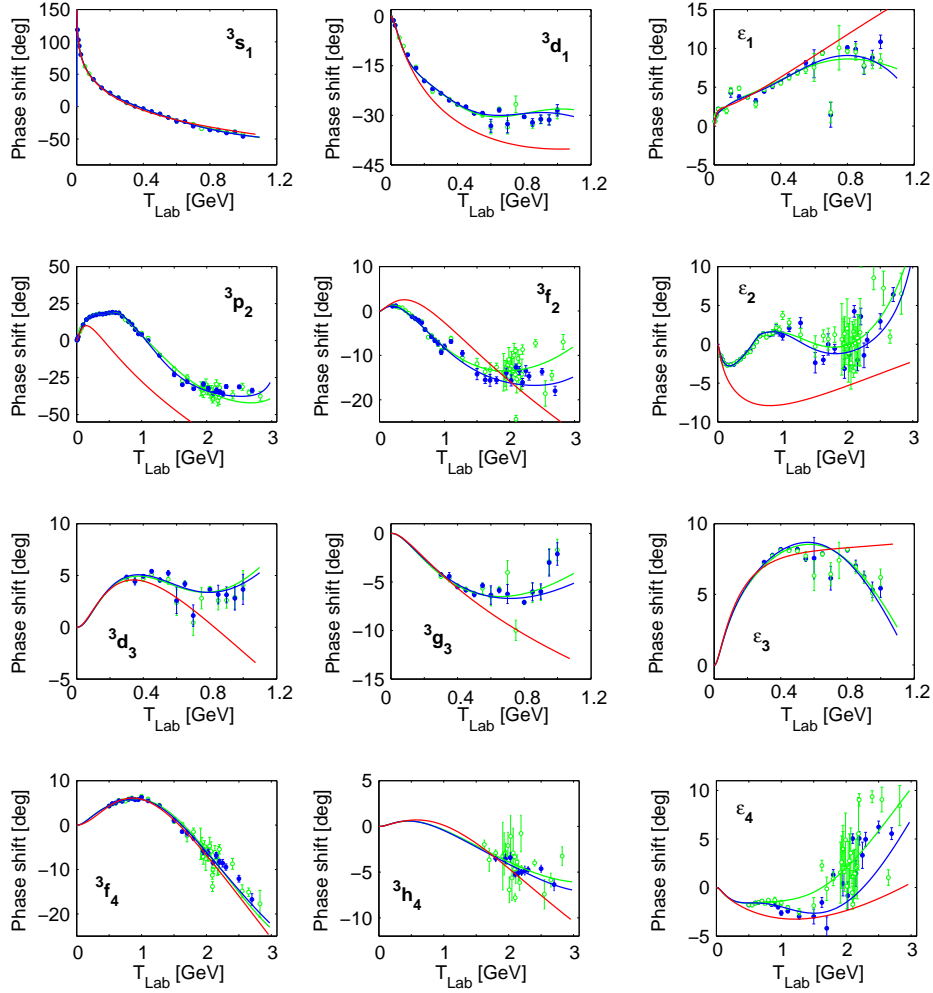}
\caption{Coupled channels
$pp$ [0,3]\,GeV, $np$ [0,1.2]\,GeV SM00 and SP03 real phase shift data $\delta(T)$ and
theoretical curve based soly upon Dirac potential (red line).
The full potential, Dirac and OMP, describes the data SP03. 
SM00 (continuous solution, green, single energy solutions, open green  circles, with error bars) 
and SP03 (continuous solutio, blue, single energy solutions,  full blue circles, with error bars)
and theoretical results (real Dirac potential solutions, red line, real Dirac potentials
with complex OMP added are {\em coinciding} with the data of SP03 blue line).
}\label{Phase_2}
\end{center}
\end{figure}

The source of the BB
channel (dibaryon formation) is the NN core-domain whose transition is mediated by a
delta-function or a narrow Gaussian function. Intrinsic nucleon excitations are 
also mediated
by a narrow Gaussian in the core domain. Inelasticities are either generated by
coupling BB (dibaryon) to asymptotic many body final states, composed of 
two nucleons and mesons, or decay of XY, composed of one or two intrinsic nucleon excitations, 
into asymptotic many body final states. Within the inner core region $r<r_c$ 
NN and XY wave function components vanish. The meson exchange  {\em Dirac potential},
which is described by the NN Dirac instant form dynamics, 
should ultimately be {\em limited to $r \ge r_c\sim 0.5\,\text{fm}$  in its  effect}. 
This constraint eliminates the need for regularization
of the short range Dirac potential and  boundary conditions
are automatically generated by the $\delta(r-r_c)$ $NN\leftrightarrow BB$ transition  potentials.
This proposal is demonstrated in Fig. \ref{Figure_5}.
The strengths and location of $\delta$-function interactions are boundary conditions which are to be
determined by BB and XY models. Herein lies the essential point of our method. Dirac potentials
play only the role of a shield which prevents us from seeing
the naked refinement surface of hadronic QCD dynamics -- it recalls the P-matrix formalism.
A realization of the full coupled channels problem is in progress.      

Without specification of details, the coupling scheme has the following structure   
\begin{eqnarray}
\mathcal{H}_{BB}\ +&\mathcal{V}_{BB}^{NN} \delta(r-r_c)&+\ \mathcal{V}_{BB} \quad\quad
\text{confined dibaryon, NN meson decay}
 \\       &\Big\updownarrow& \nonumber \\
&\mathcal{V}_{NN}^{BB}\delta(r-r_c)&+\ \mathcal{V}_{NN}^{XY}g(r-r_c)+H_{NN}+
V_{NN}^{\text{\tiny Dirac}}\quad\quad \text{NN elastic} 
\\   &  &\quad\quad\Big\updownarrow      
\nonumber \\   &  &\ \mathcal{V}_{XY}^{NN}g(r-r_c)+\mathcal{H}_{XY}+
\mathcal{V}_{XY}\quad\quad \text{NN meson decay}.
\end{eqnarray} 

Below pion production threshold, $T_{Lab}\sim 280$\,MeV, 
the NN S-matrix is unitary for all practical purposes. Above
this energy excitation of $\Delta$(3,3) resonances  is the predominant
mechanism. It  is obviously present in
the NN $^1D_2,\ ^3F_3$ and $^3PF_2$ channels. Isospin conservation suppresses
a coupling to N$\Delta$ in the  $np$ $T=0$ channels.
NN scattering, for energies
below 3 GeV in general, show a comparable to nucleon-nucleus scattering
weak and smoothly energy dependent coupling to inelastic channels. 
A perturbative treatment of inelastic and reaction channels with DWBA methods 
is thus strongly favored.

A key issue for all secondary applications of 
NN scattering is a high quality reproduction of the elastic NN scattering
channel. Inverse scattering methods are useful for this purpose.
These methods use the experimental data in form of  partial wave phase shifts as input
and determine the optical model potential as a correction 
to a theoretically defined and numerically realized  reference potential.
\clearpage  
\clearpage 
\clearpage 
\clearpage 
\clearpage 
\clearpage 
\begin{figure}[ht]
\begin{center}
\includegraphics[scale=1.0]{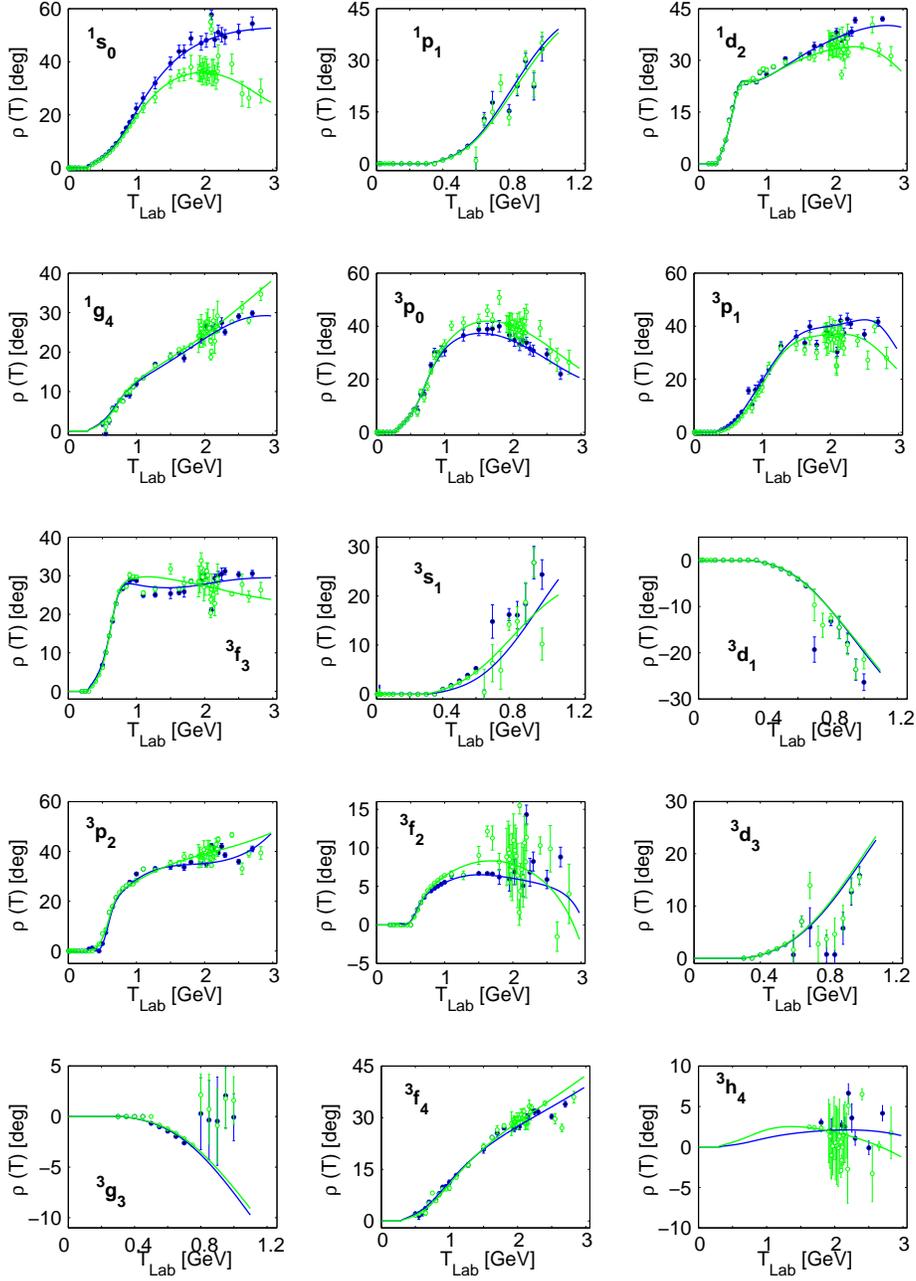}
\caption{Available absorption phase shifts $\rho(T)$. Dirac potentials
generate no absorption and  OMP parameters $U_G(T)$ and $W_G(T)$ are 
adjusted to reproduce the continuous energy solutions of $\rho(T)$. 
SM00 (continuous solution, green, single energy solutions, open green circles, with error bars) 
and SP03 (continuous solution, blue, single energy solutions, full blue circles, with error bars)
and theoretical results (real Dirac potential solutions, red, real Dirac potentials
yields no absorption but
with a complex OMP added the results are {\em coinciding} with the data of SP03 blue line).}
\label{Rho}
\end{center}
\end{figure}
\clearpage
\subsection{$^1S_0,\ ^3P_0$ and $^3P_1$ Channels}
NN phase shift data, see Figs.\,\ref{Phase_1} and \ref{Rho},
show in almost all channels and for $T_{Lab}>280$\,MeV a complicated energy dependence
and deviations from the Dirac potential predictions. 
Exceptional cases are  the  $^1S_0,\ ^3P_0$ and $^3P_1$ 
channels, contained in Fig.\,\ref{Phase_1},
 which show a {\em practical perfect reproduction}     
\begin{equation}
\delta_{Data}=\delta_{Dirac},\quad\mbox{for}\quad 0<T_{Lab}< 1100\,\text{MeV}
\end{equation} 
by the Dirac potential alone. However, the absorption 
\begin{equation}
\rho_{Data}\ne 0,\quad \mbox{whereas}\quad\rho_{Dirac}=0\quad\mbox{for}\quad 280<T_{Lab}< 1100\,
\text{MeV.}
\end{equation} 
This demands an
{\em optical potential} $U(r)+iW(r)$ that
leaves the real phase shifts $\delta(k)$ unchanged but generates an absorption
$\rho(T)>0$ for $280<T_{Lab}<1100$\,MeV.

The optical potential solutions are complex  
\begin{gather*} 
-f_r''+(V^D+U)f_r-k^2f_r=+Wf_i\\
-f_i''+(V^D+U)f_i-k^2f_i=-Wf_r
\end{gather*}
and the boundary conditions, $\delta=\delta(V^D,U,W)$ and 
$\rho=\rho(V^D,U,W)$, are transcendental functions of  potentials and 
OMP adjustable parameters.

A sensible solution is $U(r)=0$ and $W(r)=W\delta(r-r_0)$ as optical model 
interaction in Eq.\,(\ref{dirac_omp}). We used a delta-function and/or a narrow normalized
Gaussian, see Fig.\,\ref{Figure_5},
\begin{equation}
W(r)=
\begin{cases}
W_G\,N(r_0,a_0)\exp(-(r-r_0)^2/a_0^2)\\
W_{\delta}\,\delta(r-r_0)
\end{cases}
\end{equation}
with $U=0$,  starting at the origin with     
\begin{equation}
f_r(0)=0,\ f_r(h)=h^{(\ell+1)},\ f_i(r)=0\quad\mbox{for}\quad 0\le r\le r_0. 
\end{equation}
The following conclusions are drawn:
The phase-shifts $\delta(T),\,\rho(T)$ for  $280 < T_{Lab}< 1100$\ MeV
imply  an optical potential at the surface of the
repulsive core for $^1S_0,\,^3P_0$ and $^3P_1$ partial waves. Intermediate dibaryons 
are practically not formed,  the BB channel is realized by a dibaryon fusion/scission
picture as shown in Fig.\,\ref{Fusion}.
The BB dibaryon quark dynamic is reduced to energy dependent
 complex boundary conditions at the
core radius permitting meson production.
The meson exchange mechanism is not valid
inside the core radius. Caveat, at this stage of our work, we integrated from the origin through the
core region realizing a small real wave function at the core radius. This,
in connection with the Gaussian OMP form factor,  causes a real
and imaginary part OMP $U_G(T)\ne 0$ and $W_G(T)\ne 0$. 
\clearpage
\begin{figure}[ht]
\begin{center}
\includegraphics[scale=.6]{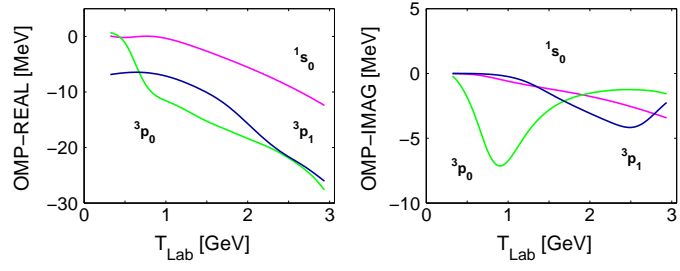}
\caption{Selected set of optical model strengths values $U_G(T)$ and $W_G(T)$.}
\label{np_OMP}
\end{center}
\end{figure}
\begin{figure}[h]
\begin{center}
\includegraphics[scale=0.25]{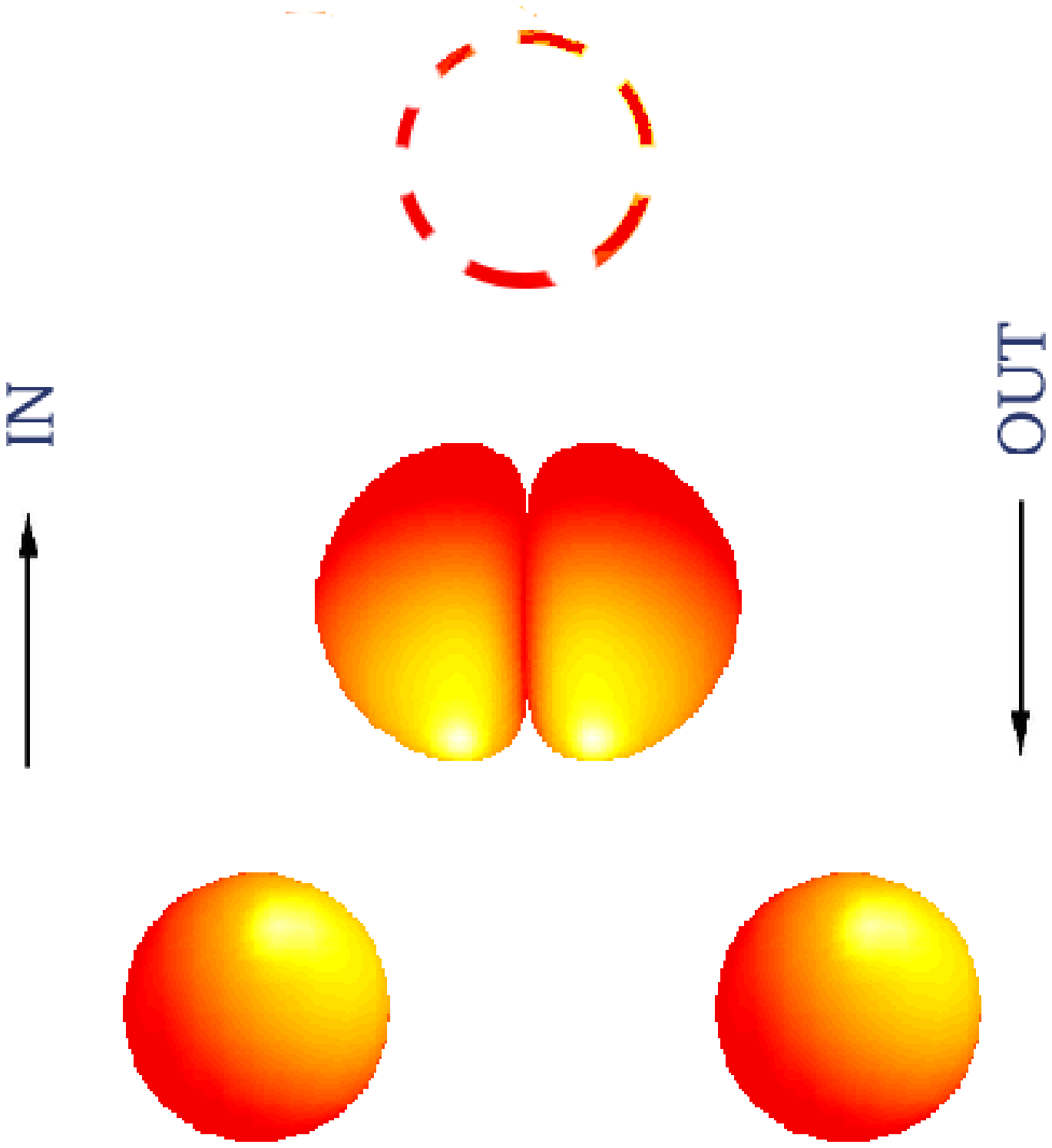}\quad
\includegraphics[scale=0.30]{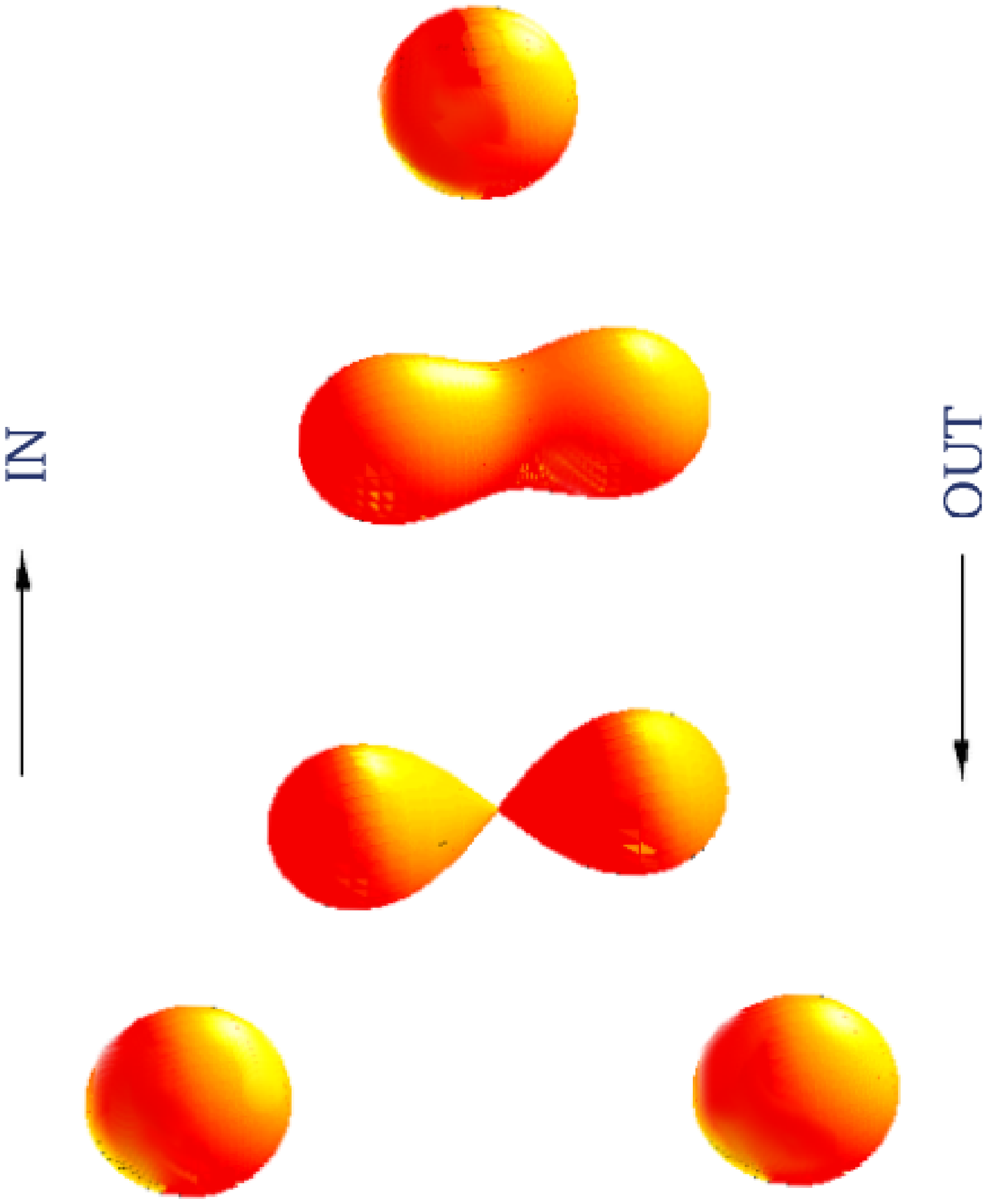}
\caption{Left, caused by the Pauli exclusion principle for a six-quark dibaryon and 
$T_{Lab,NN}<1100$\,MeV suggests a futile ringing of the nucleons and a suppression 
of dibaryon formation. 
It gives the impression of a fusion/scission mechanism.
Right, the formation of dibaryons, at sufficient high energy, is governed by medium to
longer ranged quark-gluon flux tubes with fusion into a six-quark  hadron sized dibaryon
with sequential decay. This inspires a fusion/fission mechanism.}\label{Fusion}
\end{center}
\end{figure}
In Fig.\,\ref{np_OMP} are shown the {\em np} OMP 
(normalized Gaussian, $r_0=r_c=0.5$\,fm and $a_0=0.2$\,fm) strengths values.
The crucial center of baryon $NN$ and $BB$  transition  radius is $r_0=0.5\pm0.025$\,fm.
At higher energies $T_{Lab}>1.1$\,GeV, the transition surface becomes more and more
faded, washed-out and translucent when the energy of dibaryon states matches the total
energy of the NN system. Intermediate short lived  dibaryons $J_{BB}^P= 0^+,\ 0^-,\ 1^1$  
are formed, see Fig.\,\ref{Fusion}.
\\[0.1cm]
We estimate, from the phase behavior in these three channels, the total energy (lowest mass) 
of a dibaryon system   $m_{BB}= 2400\pm 150$\,MeV and a  width $\Gamma> 150$\,MeV.  
\\[0.1cm] 
Coupling to dibaryons is realized  for $T_{Lab}>1100$\,MeV and the fusion/scission picture
may change gradually into a fusion/fission picture as shown in Fig.\,\ref{Fusion}.
\section*{Acknowledgment}
The authors wish to thank H. Crater and B. Liu for valuable correspondence.
One of us, DB appreciates the support by the DAAD.

\end{document}